\documentclass{iau}

\usepackage{amsmath}
\usepackage{graphicx}
\usepackage{multirow}
\usepackage{natbib}

\begin{document}

\lefttitle{S. Sim\'on-D\'iaz}
\righttitle{How large spectroscopic surveys are shaping our understanding of massive stars}

\jnlPage{1}{7}
\jnlDoiYr{2026}
\doival{}

\aopheadtitle{Proceedings IAU Symposium}
\editors{A. Wofford, L. Oskinova, M. Garcia, N. St-Louis, S. Sim\'on-D\'iaz, eds.}

\title{How large spectroscopic surveys are shaping \\ our understanding of massive stars}

\author{S. Sim\'on-D\'iaz}
\affiliation{Instituto de Astrof\'isica de Canarias \& Universidad de La Laguna}

\begin{abstract}
We present an overview of the main characteristics of several spectroscopic surveys designed to advance our understanding of the physical properties and evolution of massive stars. We also summarize key results obtained from the analysis of these datasets, highlighting how the interpretation of some observables in the framework of massive-star evolution is considerably more complex than previously anticipated.
\end{abstract}

\begin{keywords}
Massive OB stars, spectroscopic observations, surveys, stellar parameters (incl. spin rates and winds), surface abundances, spectroscopic binaries, stellar evolution.
\end{keywords}

\maketitle

\section{Introduction}

It is no coincidence that, for the first time, an IAU Massive Star Symposium includes a dedicated session on spectroscopic surveys. For more than two decades, the community has been assembling high-quality spectra of increasingly large samples of massive stars in several galaxies within the Local Volume. What began modestly in the first decade of the 2000s, with the design and early scientific results of initiatives such as the {\em VLT-FLAMES Survey of Massive Stars}, the {\em Galactic O-Star Spectroscopic Survey}, or the {\em MiMeS} and {\em IACOB} projects, 
soon evolved into a central observational strategy in massive-star research. With more than a dozen purpose-built spectroscopic surveys -- mostly focused on O- and B-type stars -- in the Milky Way and the Magellanic Clouds,
some of them completed, others still ongoing, and a few more about to begin,
there is now a wealth of results to be presented and discussed.

While many of these results, together with detailed descriptions of the associated surveys, are presented in subsequent contributions to these proceedings, we provide here a personal perspective on how these community-driven observational efforts have, over time, shaped our understanding of the physical properties and evolution of massive stars. 

\section{Key ingredients of a successful spectroscopic survey}\label{ingredients}

Before addressing the main topic of this contribution, we summarize several key ingredients that must be carefully considered to ensure the scientific success of a spectroscopic survey: 

\begin{itemize}
  \item {\bf Scientific scope:} This first ingredient may appear obvious at first glance, yet it constitutes a fundamental pillar of any spectroscopic survey. It involves the clear identification of the key scientific question(s) to be addressed, together with a careful assessment of the type of information required to answer them. In particular, this may include the identification of previously unknown massive-star populations and their spectral classification; kinematic studies; the determination of physical parameters and/or surface abundances; the detection and characterization of spectroscopic binaries -- including their orbital properties -- as well as stellar variability and/or magnetic fields, among other relevant observational diagnostics.
  \item {\bf Target selection:} This ingredient is naturally connected to the previous one and involves the definition of the types of stars to be targeted, the regions of the Galaxy or external galaxies on which the survey will focus, and the evaluation of the minimum, optimal, and ideal sample sizes required to ensure that the scientific questions can be properly addressed. In this context, it is also essential -- particularly when statistical robustness is required or when observational results are to be compared with theoretical predictions -- to assess whether the surveyed sample may be affected by significant observational biases. Finally, it is important to consider which sources of existing information (e.g., published catalogs, photometric data) are available, and which candidate-selection criteria will be applied to construct the target list.
  \item {\bf Observational strategy:} Acquiring the observations required to address specific scientific questions demands appropriate telescope facilities and instrumental capabilities. These include telescope aperture; wavelength coverage (optical, infrared, or ultraviolet, in some cases requiring space-based facilities); spectral resolution; and the ability to achieve the signal-to-noise ratios needed for the selected spectral resolution and magnitude range of the targeted stellar sample. Observational efficiency can be further enhanced through the use of multiplexing capabilities -- such as multi-object spectroscopy (MOS) or integral-field spectroscopy (IFS/IFU) -- rather than single-fibre or single-slit instruments. In addition, the value of obtaining multi-epoch observations, together with the identification of an optimal cadence, total time span, and number of epochs, must also be carefully considered.
  \item {\bf Data analysis:} Beyond the compilation of observations -- and despite the fact that most observatories now provide data in formats that are close to being ready for scientific analysis -- there remain several steps that require special attention in order to efficiently translate observed spectra into scientifically useful quantities, particularly in the context of large surveys. The first concerns the filtering, classification, and pre-processing of the data, especially for surveys that employ multiplexing capabilities and can deliver spectra for several hundred, or even thousands of targets in a single observing run. The second involves the more computationally demanding quantitative spectroscopic analysis itself. In this context, and in view of the steadily increasing volume of data to be analysed, fast (semi-)automated analysis tools -- often incorporating advanced computational techniques based on Bayesian inference and various machine-learning approaches -- are being increasingly developed and adopted.
  \item {\bf Additional aspects:} In addition to the abovementioned (main) ingredients, experience over the past decades has highlighted the importance of several additional aspects. Firstly; while spectroscopic surveys are normally led by observers, their scientific impact is significantly enhanced when the team includes several theoreticians. Secondly, their impact is further strengthened when complemented by additional observations\footnote{See notes on the interest of astrometry, time-domain photometry, and interferometry in massive star research in the contributions to these proceedings by Maíz Apellániz, Pedersen, Borges, and Sana.}, and through the development of synergies with other projects. Last but not least, it is highly desirable to devote effort to optimizing the long-term scientific legacy of a survey. This includes not only the dissemination of results through refereed publications and conference presentations, but also the timely public release of the compiled observations and analysis products -- ideally through well-documented online databases -- so that they can be  accessed by the wider community.
\end{itemize}

\section{Past and on-going spectroscopic surveys of massive stars in a nutshell}\label{surveys}

\subsection{Surveys (mostly in the Milky Way) based on single-object observations}\label{singleobject}

The {\bf \em Galactic O-Star Spectroscopic Survey} \citep[GOSSS;][]{MaizApellaniz2011, MaizApellaniz2017} was initiated in 2006 with the ambitious goal of obtaining modern blue–violet spectra (R $\approx$ 2500) of all Galactic stars previously classified as O–B0, thereby supporting and extending the Galactic O-Star Catalog \citep[GOSC;][]{MaizApellaniz2013}. After almost two decades of regular single-object observations, primarily using medium-sized telescopes in both the Northern and Southern hemispheres, GOSSS has reached the milestone of collecting at least one spectrum for more than 1000 Galactic O-type stars. 
This survey has become a key reference for the community as a result of two major achievements: (1) the establishment of a homogeneous and widely adopted reference framework for O-star spectral classification, including a revision of previously proposed criteria and the construction of a low-resolution spectral atlas of standard stars \citep{MaizApellaniz2015}; and (2) a comprehensive revision of the spectral classifications of all stars currently listed in GOSC, which can be regarded as the most complete census of known Galactic O-type stars \citep{Sota2011, Sota2014, MaizApellaniz2016}. This later achievement includes the identification of stars exhibiting unusual spectral properties (e.g. Oe and Of?p objects) as well numerous spectroscopic binary systems.

GOSSS is currently transitioning into the Alma Luminous Star (ALS) survey. Following a similar observing strategy, Maíz Apellániz aims to expanding the targeted sample to encompass all types of blue luminous stars \citep[e.g.][]{Pantaleoni2025} and extending the survey coverage to include not only the Milky Way, but also both Magellanic Clouds.


The {\bf \em OWN survey} \citep[][]{Gamen2007} also started in 2006 with the primary goal of characterizing the multiplicity properties of southern Galactic O- and WN-type stars, and of determining the orbital parameters of those systems identified as spectroscopic binaries. To this end, the survey has carried out systematic multi-epoch, single-object, high-resolution optical spectroscopic observations over nearly two decades, primarily using 2-m class telescopes in Argentina and Chile.
As a result, OWN has collected approximately 6\,000 high-quality spectra for a sample of about 200 O-type stars and 60 WN stars \citep{Barba2017}. As described by Morrell et al. (these proceedings; see also Gamen et al., in prep.), in addition to classifying the full sample into likely single stars, radial-velocity variables, and spectroscopic binaries, one of the most significant achievements of the OWN project has been the determination of high-precision radial-velocity orbits for 41 SB2 and 30 SB1 systems. Part of the observations gathered by the OWN survey -- particularly those obtained with the FEROS instrument -- have also proven valuable for other projects, such as IACOB (see below and Sect.~\ref{highlights}), as well as for a number of targeted studies \citep[e.g.,][]{Martins2015b, Martins2017, Burssens2020}.

The {\bf \em IACOB survey} \citep{SimonDiaz2011a, SimonDiaz2020} has, since 2008, been intensively gathering high-resolution (R\,=\,25000\,--\,85000) spectra in the range between $\sim$380 and 900~nm for all bright (V$\lesssim$9~mag) Galactic OB-type stars observable with the NOT (2.56\,m) and Mercator (1.2\,m) telescopes operating at the Roque de los Muchachos Observatory in the Canary Islands. The survey was motivated by the goal of generating a comprehensive catalog of observational properties for a statistically significant sample of blue massive stars, intended to serve as a robust and long-lasting anchor for theoretical studies of stellar atmospheres, winds, internal structure, and the evolution of Galactic massive stars. This continuously increasing catalog primarily comprises stellar parameters and surface abundances \citep[e.g.,][]{deBurgos2024, Holgado2025, SimonDiaz2026}, but also spectroscopic variability and binarity \citep[e.g.,][]{SimonDiaz2017, SimonDiaz2024}.  After benefiting from spectra of comparable quality for southern OB-type stars obtained with the FEROS instrument and available in the ESO archives\footnote{Including a combination of spectra obtained by the OWN survey and several other independent observing programmes.}, the associated IACOB spectroscopic database presently comprises approximately 15\,000 spectra for about 2\,800 stars. Several interesting scientific results obtained from the analysis of this unique spectroscopic dataset are presented in Sect.~\ref{highlights}.

The {\bf \em MiMeS survey} (Magnetism in Massive Stars; \citealt{Wade2016}) focused on the acquisition of more than 4\,800 very high signal-to-noise ratio, high-resolution (R$\gtrsim$65000), circularly polarised (Stokes $I$ and $V$) optical spectra of several hundred Galactic stars with spectral types ranging from O4 to B9.5. The observations were obtained between 2008 and 2013 using several high-resolution spectropolarimeters mounted on 3–4 m class telescopes in both hemispheres. The survey was designed to enable the first large-scale, systematic investigation of magnetism in massive stars.
In addition to a dedicated Targeted Component, which obtained time-domain observations of approximately 30 known or suspected magnetic hot stars \citep[see the reference list of targets and associated publications in][]{Wade2016}, the main Survey Component comprised about 106 O- or WR-type stars and 422 B-type stars, with V$\lesssim$13.6~mag. Among the key results derived from the analysis of the Survey Component, MiMeS established, for the first time, a statistically significant incidence rate of detectable magnetic fields among O-type stars \citep[7$\pm$3\%,][]{Grunhut2017},
found no clear link between stellar physical properties and the presence of a magnetic field, and suggested that the initial magnetic-field distribution is likely bimodal, with young O-type stars hosting either weak/absent or strong magnetic fields \citep{Petit2019}. Furthermore, quantitative spectroscopic analyses of O-type stars in the MiMeS sample not identified as SB2 enabled the first large-scale investigation of CNO surface abundances in Galactic O-type stars \citep{Martins2015a}. 

Two closely related spin-off projects emerged from MiMeS: {\bf \em Binarity and Magnetic Interactions in various classes of stars} (BinaMIcS; \citealt{Alecian2015}) and {\bf \em Magnetic OB[A] Stars with TESS: probing their Evolutionary and Rotational properties} (MOBSTER; \citealt{DavidUraz2019}). Also worth mentioning in this context is the {\bf \em B Fields in OB Stars} Survey (BOB; \citealt{Morel2015}). Together, these initiatives have provided complementary constraints on the incidence, origin, and properties of magnetic fields in massive stars.

The {\bf \em ULLYSES} (UV Legacy Library of Young Stars as Essential Standards; \citealt{RomanDuval2020}) and {\bf \em XShootU} \citep{Vink2023} programmes represent two closely connected initiatives. Active between 2019 and 2023, ULLYSES was a major DDT Program on the Hubble Space Telescope (HST), designed to create a UV spectroscopic legacy dataset of both high- and low-mass stars in the local Universe. It represents one of the largest HST programmes ever executed, using approximately 1\,000 orbits. The massive-star component of ULLYSES comprises about 250 OB stars in the LMC, SMC, Sextans A, and NGC~3109, sampling a wide range of metallicities from about 0.5 to 0.1 solar. This dataset is complemented by medium-resolution (R$\sim$7\,000\,--\,12\,000) spectroscopy of a similar target sample, covering wavelengths from the near-UV ($\gtrsim$\,300~nm) to the near-IR ($\lesssim$\,2500~nm), obtained with the X-shooter instrument at the VLT (8.2~m) thanks to a $\sim$126~h awarded ESO Large Programme. Through a coordinated community effort, the analysis of these complementary datasets has enabled significant progress on several long-standing questions in massive-star research \citep[see][and Tramper, these proceedings]{Vink2024}, most notably regarding the fundamental properties of radiatively driven winds in massive OB-type stars and their dependence on metallicity \citep[e.g.,][among many others]{BerniniPeron2024, Hawcroft2024, Brands2025}.
%
\subsection{Multi-object spectroscopic (MOS) surveys in the Milky Way and the Magellanic Clouds}\label{mos}
The {\bf \em VLT-FLAMES Survey of Massive Stars} (VFSMS; \citealt{Evans2005}) was the first large-scale project dedicated to massive stars to take full advantage of a MOS facility. The survey was designed to obtain high-quality optical spectra of a statistically significant, magnitude-limited sample of O- and B-type stars in several stellar clusters across the Milky Way (MW) and both Magellanic Clouds (MCs), with the primary goals of testing the efficiency of rotational mixing, and investigating the dependence of stellar-wind properties on metallicity.
Based on a specifically designed ESO Large Programme using FLAMES at the VLT, VFSMS obtained high-resolution (R$\sim$25\,000) spectra in two optical wavelength ranges (385\,--\,475 and 638\,--\,662~nm) for approximately 50 O-type stars (spanning all luminosity classes) and about 500 B-type stars (predominantly dwarfs and giants). Targets were homogeneously selected from photometry in three MW clusters ($Z\sim$\,$Z_{\odot}$) and two clusters in each of the MCs (LMC, $\sim$\,0.5\,$Z_{\odot}$; SMC, $\sim$\,0.2\,$Z_{\odot}$). As in the case of the MiMeS survey, VFSMS was supported from its outset by a medium-sized international collaboration combining observers with experts in the modelling of stellar atmospheres and stellar evolution. This close interaction enabled a more efficient and physically grounded interpretation of the observational results. As discussed in Sect.~\ref{highlights}, the survey yielded several important outcomes, including results that challenged aspects of stellar evolution theory previously regarded as well established.

The {\bf \em VLT-FLAMES Tarantula Survey} (VFTS; \citealt{Evans2011}) was a natural follow-up to VFSMS. Building on the success of the earlier programme and motivated by the challenges it posed to existing theoretical frameworks, members of the same collaboration designed a new survey making again use of the FLAMES instrument. This time the primary goal was to carry out a comprehensive investigation of the massive-star population in the 30 Doradus region of the LMC. The survey aimed to obtain homogeneous estimates of stellar parameters and surface abundances and, crucially, to quantify for the first time the incidence and properties of massive binary systems in this nearby low-metallicity starburst environment. Between 2008 and 2011, VFTS obtained high-quality spectra covering the wavelength ranges 396\,--\,570 and 644\,--\,687 nm at spectral resolutions of R$\sim$7\,500 and $\sim$16\,000, respectively, for more than 800 massive O, B, and WR stars across the 30 Doradus complex. In particular, six-epoch observations in the 396–456 nm range were secured to enable the efficient identification of spectroscopic binaries. Five additional pointings centred on the core of 30 Doradus (the R136 cluster) were obtained with the ARGUS IFU. Some key scientific results derived from VFTS are presented in Sect.~\ref{highlights}. 

Two spin-off surveys performing additional longer-term multi-epoch observations of 93 O-type and 88 early B-type stars identified as spectroscopic binaries by VFTS are the {\bf \em Tarantula Massive Binary Monitoring} \citep[TMBM,][]{Almeida2017} and the {\bf \em B-type binaries characterization programme} \citep[BBC,][]{Villasenor2021}, respectively.

For completeness, while not described here in detail, as they are presented elsewhere in these proceedings by Villaseñor and Vargas-Salazar, respectively, is worth mentioning the {\bf \em Binarity at LOw Metallicity} campaign (BLOeM; \citealt{Shenar2024}, a natural continuation of VFTS) and the {\bf \em Runaways and Isolated O-Type Star Spectroscopic Survey of the SMC} (RIOTS4; \citealt{Lamb2016}). Led by independent teams, both surveys cover the full star-forming body of the SMC and are based on a long-term (three and five years, respectively), medium-resolution (R\,$\sim$7000 and 3000, respectively) spectroscopic monitoring of approximately 900 and 300 targets, respectively, in this low-metallicity galaxy. Also, the {\bf \em Gaia–ESO Survey} (GES; \citealt{Gilmore2022, Randich2022}), a large community-driven effort, included a dedicated, though relatively small, component focused on the massive-star populations of several star-forming regions and open clusters in the Milky Way \citep{Blomme2022}. To date, the most relevant results concerning OB-type stars obtained within the framework of this survey have primarily focused on the Carina region \citep[e.g.,][]{Berlanas2023, Berlanas2025, Santos2025}.

\section{How large spectroscopic surveys are shaping massive star research}\label{highlights}

The first IAU symposium dedicated to massive stars (IAUS\#49, Buenos Aires, Argentina, 1971) was devoted to Wolf–Rayet and High-Temperature Stars. The second one (IAUS\#83, Qualicum Beach, Canada, 1978) focused on Mass Loss and Evolution of O-Type Stars. As poignantly recalled by Virpi Niemela during the conference held in her honour in Cariló (Argentina) in 2007, {\em “[by that time] we had just learned from UV observations that mass loss was profoundly affecting the evolution of all massive stars, not only a few”}. 
Following these first two gatherings, a succession of additional massive-star “Beach Symposia” took place at roughly 4-year intervals,
reflecting the rapid pace of progress in the field.

For the two decades following IAUS\#83, observational and theoretical advances achieved by a steadily growing community were largely driven by a gradual -- but heterogeneous -- accumulation of spectroscopic observations of OB, WR, and red supergiant stars, assembled using a wide variety of instruments, observing strategies, and analysis techniques. Although sample sizes in most observational studies only rarely exceeded a few tens of objects, by the turn of the millennium the progress achieved in both the theory of radiatively driven winds and the modelling of massive-star evolution had been substantial, as comprehensively reviewed by \citet{KudritzkiPuls2000} and \citet{MaederMeynet2000}.

It was only in 2003, with the advent of the VLT-FLAMES Survey of Massive Stars, that a new era in massive-star research truly began: the era of large spectroscopic surveys. As described in Sect.~\ref{surveys}, several other initiatives with similar observational philosophies -- though focusing on different scientific questions, stellar samples, and metallicity environments -- soon followed, many of them operating in parallel. One immediate consequence was a rapid and sustained increase in the number of available high-quality spectra, reaching several hundred stars in most cases.
This development, in turn, required a paradigm shift in the way such data were processed and analysed. In parallel with the availability of a new generation of stellar atmosphere codes (see reviews by \citealt{Puls2009} and \citealt{Sander2017}), the growing size of the datasets necessitated the development of automated quantitative spectroscopic analysis techniques, enabling a more efficient and objective determination of stellar parameters and surface abundances \citep[e.g.,][see also the contribution by Crowther \& Bestenlehner to these proceedings]{Mokiem2005, SimonDiaz2011b}. Strengthening the collaboration between observers and modelers also became increasingly important in order to develop optimal strategies for comparing observational results with the predictions of theoretical models.

The VLT-FLAMES Survey of Massive Stars enabled the first statistically significant confrontation between observations and theoretical predictions regarding the dependence of stellar-wind properties on luminosity and metallicity in O- and early B-type stars, yielding encouraging results \citep{Mokiem2007}. In addition, the survey confirmed long-standing expectations that main-sequence massive stars at low metallicity exhibit, on average, higher rotational velocities \citep{Hunter2008b}. However, it also  
challenged rotational mixing as the dominant mechanism responsible for the presence of massive main-sequence stars with surface abundances enriched in CNO-cycle products \citep{Hunter2008a, Brott2011}.

By the start of the new millennium, a broad consensus had emerged that rotation and stellar winds -- alongside stellar mass -- were the dominant agents governing the evolution of massive stars. In this context, rotational mixing was proposed as a plausible explanation for the presence of stellar surfaces enriched in CNO-cycle products. This mechanism can lead to a significant redistribution of chemical elements during the core hydrogen-burning phase. As a consequence, a clear correlation between surface nitrogen abundances and projected rotational velocities ($v$\,sin$i$) was expected. This prediction was put to the test when members of the VFSMS collaboration carried out a detailed spectroscopic analysis of a sample of 135 early B-type stars in the LMC, spanning projected rotational velocities up to $\sim$300\,km\,s$^{-1}$. The analysis, first presented in December 2007 at the IAUS\#250 (Kauai, USA), revealed a significant population ($\approx$20\% of the sample) of highly nitrogen-enriched, intrinsically slow rotators ($v$\,sin$i\lesssim$50\,km\,s$^{-1}$). Furthermore, an additional 20\% of the sample consisted of rapidly rotating stars that exhibited little surface nitrogen enrichment. Both results appeared to challenge the concept of rotational mixing, and urgently advocate for additional explanations (including binaries, magnetic fields and internal gravity waves). 

This was not the first time that a similar result had been reported \citep{Morel2008}; however, the availability of a larger and more homogeneous sample significantly increased the statistical robustness of the finding. In addition, the possibility of directly comparing the observational results with stellar-evolution models computed specifically for that study \citep{Brott2011} proved crucial for identifying the regimes in which alternative physical processes or additional mechanisms needed to be explored. Among these, one factor that had been repeatedly emphasized over many years -- most notably by D. Vanbeveren in successive IAU Massive Star Symposia since IAUS\#83 -- was the role of binarity in the evolution of massive stars.

The VLT-FLAMES Tarantula Survey provides a particularly clear example of how a large spectroscopic survey can efficiently advance the field. Among the many important results obtained from the thorough investigation of the massive-star population of 30 Doradus \citep{Evans2020}, one of the most significant outcomes was the demonstration that more than 50\% of the O- and B-type stars in this starburst-like region are members of binary or higher-order multiple systems \citep{Sana2013, Dunstall2015}. These results, together with findings from parallel contemporary investigations \citep[summarized at IAUS\#329 in Auckland, Australia by][]{Sana2017}, led to the definitive confirmation that binary interaction plays a dominant role in the evolution of massive stars \citep{Sana2012}.

Multi-epoch spectroscopic surveys focused on the detection and detailed orbital characterization of massive binary systems -- such as OWN, BLOeM, and RIOTS4, among others (see Sect.~\ref{surveys}) -- have since become critical for establishing the role of binarity in massive-star evolution. When combined with modern calculations of binary stellar evolution \citep{Langer2012, Marchant2024}, they provide a solid observational and theoretical foundation for assessing the expected incidence of binary interaction on the life cycles of massive stars. 

However, in light of several additional insights that have emerged over the past decade in the context outlined above \citep[e.g.,][]{deMink2013, deMink2014, Sana2022, Shenar2022, Mahy2022, Menon2024}, this is not the only observational challenge that must be addressed to ensure continued progress in the field. In particular, caution is required when interpreting results derived from the spectroscopic analysis of populations of apparently single stars, as treating them as objects that were born genuinely single can lead to spurious conclusions. Furthermore, a comprehensive understanding of the impact of binarity on massive-star evolution also requires an in-depth investigation of the observational properties of large populations of apparently single stars, as well as systems identified as SB1 binaries.
This follows from the realization that a significant fraction of the products of massive binary interaction may be observationally classified as apparently single stars or SB1 systems, depending on the nature of the interaction and/or on whether the binary has been disrupted following the supernova explosion of the initially more massive companion.

In this context, several recent studies based on the analysis of spectra compiled in the IACOB spectroscopic database have focused on identifying observational diagnostics that allow the disentanglement of massive binary-interaction products (BIPs) from stars that have evolved genuinely as single objects since birth. As representative examples, \citet{Holgado2022} and \citet{Britavskiy2023} evaluated the hypothesis proposed by \citet{deMink2013} that the high-velocity tail of the projected rotational velocity distribution of O-type stars is predominantly populated by BIPs.
By combining information on stellar parameters (including rotational velocities), radial velocities, and the runaway nature of a sample of several hundred Galactic O-type stars, these authors found observational evidence indicating that single O-type stars are most likely born with rotational velocities below $\sim$20\% of their critical velocity, while the population of fast rotators is largely composed of mass gainers. A significant fraction of these fast rotators was also identified as runaway stars \citep[see also][]{CarreteroCastrillo2025}.
More recently, \citet{MartinezSebastian2025} identified a subpopulation of Galactic O-type stars -- amounting to approximately 20\% of a sample of 180 stars with $v$\,sin$i<$150~km\,s$^{-1}$) -- exhibiting a combined He and N surface abundance pattern that cannot be explained by internal mixing in single-star evolutionary models. In addition, \cite{SimonDiaz2026} have proposed the surface He abundance as a robust proxy for identifying BIPs among O-type stars.

\section{Concluding remarks and future prospects}\label{surveys-future}
Three main take-home messages emerge, in my view, from the work carried out to date using spectroscopic observations from large surveys.
\begin{itemize}
\item Contrary to early expectations, analyses of high-quality -- often multi-epoch -- spectra from medium- and large-scale surveys have challenged our previous textbook paradigm of massive star evolution. This outcome reflects the intrinsic complexity of the problem: the evolution of massive stars depends on the interplay between mass, rotation, stellar winds, and binary interaction, together with an intricate coupling among multiple physical processes.
\item Despite some progress from the modelling side, our current understanding of massive star evolution resembles an incomplete jigsaw puzzle, in which several key pieces are still missing. In this context, the reference “cover image” provided by existing evolutionary models may be incomplete or, in some cases, misleading. Moreover, some puzzle pieces may not correspond to what was initially assumed -- for example, we might be erroneously classifying some binary interaction products as genuinely single stars -- while others may not yet have been correctly incorporated into the picture, such as runaway stars.
\item 
We are currently in the midst of a transformative era in which large, purpose-designed surveys of massive stars play a central role. Making progress in reassembling the jigsaw puzzle of massive star evolution requires the combination of as many complementary observational diagnostics as possible, including stellar parameters, surface abundances, as well as dynamical and environmental information. In this respect, time-domain photometry -- enabling the identification of eclipsing systems and asteroseismic investigations -- as well as high-angular-resolution imaging and interferometry -- providing sensitivity to close companions -- represent particularly valuable additional pieces.
\end{itemize}

We are confident that, at the next IAU Symposium on massive stars, a dedicated session on spectroscopic surveys will again be needed to provide an updated overview of the advances in the field enabled by the ongoing efforts of several groups worldwide. This enterprise will benefit not only from the observations provided by the spectroscopic surveys discussed in these proceedings, but also from several ongoing and forthcoming, even larger-scale surveys in the Milky Way and the Magellanic Clouds that include a substantial massive star component. Among these, spectroscopic observations obtained in the context of the {\em LAMOST Experiment for Galactic Understanding and Exploration} \citep[LEGUE,][]{Deng2012} and the {\em SDSS-V/Milky Way Mapper} \citep[MWM, e.g.,][]{Zari2021} programmes will provide valuable contributions at a resolving power of R$\sim$1800 for Galactic OB stars. Most importantly, the WEAVE-SCIP \citep{Jin2024}, 4MIDABLE-LR \citep{Chiappini2019}, and 1001MC \citep{Cioni2019} surveys are expected to be transformative, enabling an increase of more than one order of magnitude in the number of massive stars in the Milky Way and the Magellanic Clouds for which optical, mid-resolution (R$\sim$5000) spectra will be available.

\end{document}